 \documentclass[final,1p,times]{elsarticle} 

\journal{Nuclear Physics A} 

\usepackage{graphicx}
\usepackage{amssymb} 
\usepackage{amsthm} 
\usepackage{lineno}

\begin{document}
\begin{frontmatter} 
\title{Elliptic flow of J/$\psi$ at forward rapidity in Pb-Pb collisions at 2.76~TeV with the ALICE experiment}

\author{Hongyan Yang (for the ALICE\fnref{col1} Collaboration)}
\fntext[col1] {A list of members of the ALICE Collaboration and acknowledgements can be found at the end of this issue.}
\address{SPhN/Irfu, CEA-Saclay, Orme des Merisier, 91191 Gif-sur-Yvette, France}

\begin{abstract} 
We present the elliptic flow of inclusive J/$\psi$ measured in the $\mu^{+}\mu^{-}$ channel at forward rapidity ($2.5<y<4.0$), down to zero transverse momentum, in Pb-Pb collisions at $\sqrt{s_{_{\mathrm{NN}}}}=2.76$~TeV with the ALICE muon spectrometer. The $p_{\mathrm T}$ dependence of J/$\psi$ $v_2$ in non-central (20\%-60\%) Pb-Pb collisions at $\sqrt{s_{_{\mathrm{NN}}}}=2.76$~TeV is compared with existing measurements at RHIC and theoretical calculations. The centrality dependence of the $p_{\mathrm{T}}$-integrated elliptic flow, as well as the $p_{\mathrm{T}}$ dependence in several finer centrality classes is presented. 
\end{abstract} 

\end{frontmatter} 

\linenumbers

\section{Introduction}
\label{intro}
Charmonium production in heavy ion collisions has been studied at different energies and with different collision systems, ever since the J/$\psi$  suppression induced by color screening of its constituent quarks was proposed as a signature of the formation of a quark gluon plasma (QGP) in heavy-ion collisions~\cite{satz}. 
The recent measurement of the J/$\psi$ production in Pb-Pb collisions at forward rapidity performed by ALICE at the LHC~\cite{aliceJpsiPRL} clearly showed less suppression compared with SPS and RHIC results~\cite{spsJpsi, phenixJpsi}. 
At RHIC energies, the preliminary result from the STAR collaboration showed a J/$\psi$ elliptic flow in Au-Au collisions at  $\sqrt{s_{_{\mathrm{NN}}}}=200$~GeV~\cite{starJpsiFlow} consistent with zero within uncertainties in the measured $p_{\mathrm{T}}$ range (0-10~GeV/$c$). The measurement of quarkonium elliptic flow is especially promising at the Large Hadron Collider (LHC) where the high energy density of the medium and the large number of $c\bar{c}$ pairs produced in Pb-Pb collisions is expected to favor the flow development and regeneration scenarios.

\section{Data analysis and results}
\label{analysis}
The ALICE detector is described in~\cite{aliceJINST}. At forward rapidity ($2.5 < y < 4.0$) the production of quarkonium states is measured in the muon spectrometer down to $p_{\mathrm{T}}=0$. 
The data sample used for this analysis corresponds to 17~M dimuon unlike sign (MU) triggered Pb-Pb collisions collected in 2011. 
It corresponds to an integrated luminosity $\it{L}_{\mathrm{int}} \approx {\rm 70}~\mu {\mathrm{b}}^{\mathrm{-1}}$. 
The event and muon track selection are the same as described in~\cite{qmJpsiTalk}, except for an additional requirement of the event vertex position $|Z_{\mathrm {vtx}}|<10$~cm to ensure a flat event plane distribution. J/$\psi$ candidates are formed by combining pairs of opposite-sign (OS) tracks reconstructed in the geometrical acceptance of the muon spectrometer. 

\begin{figure}[htbp]
\begin{center}
\includegraphics[width=0.49\linewidth,keepaspectratio]{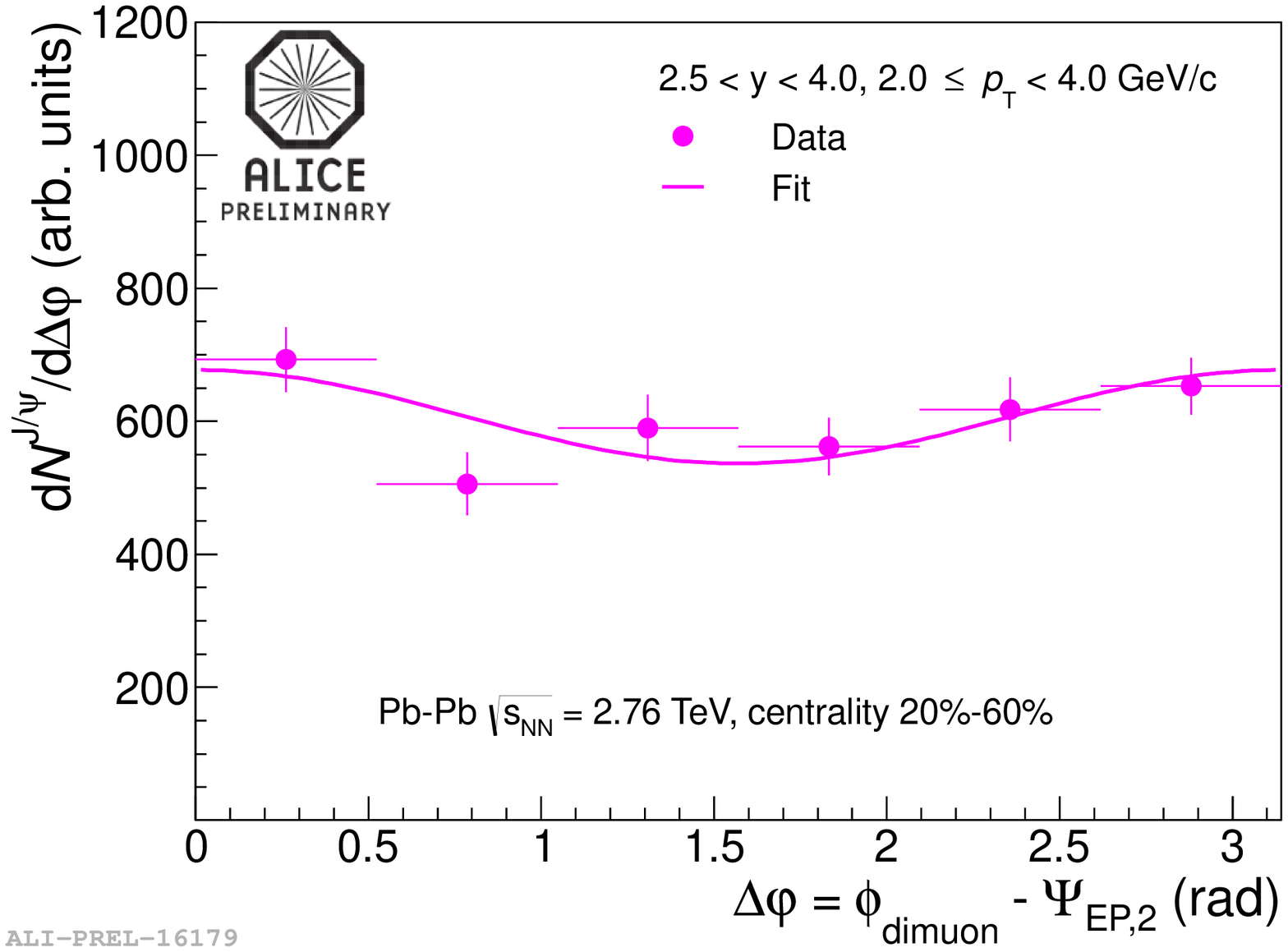} 
\includegraphics[width=0.49\linewidth,keepaspectratio]{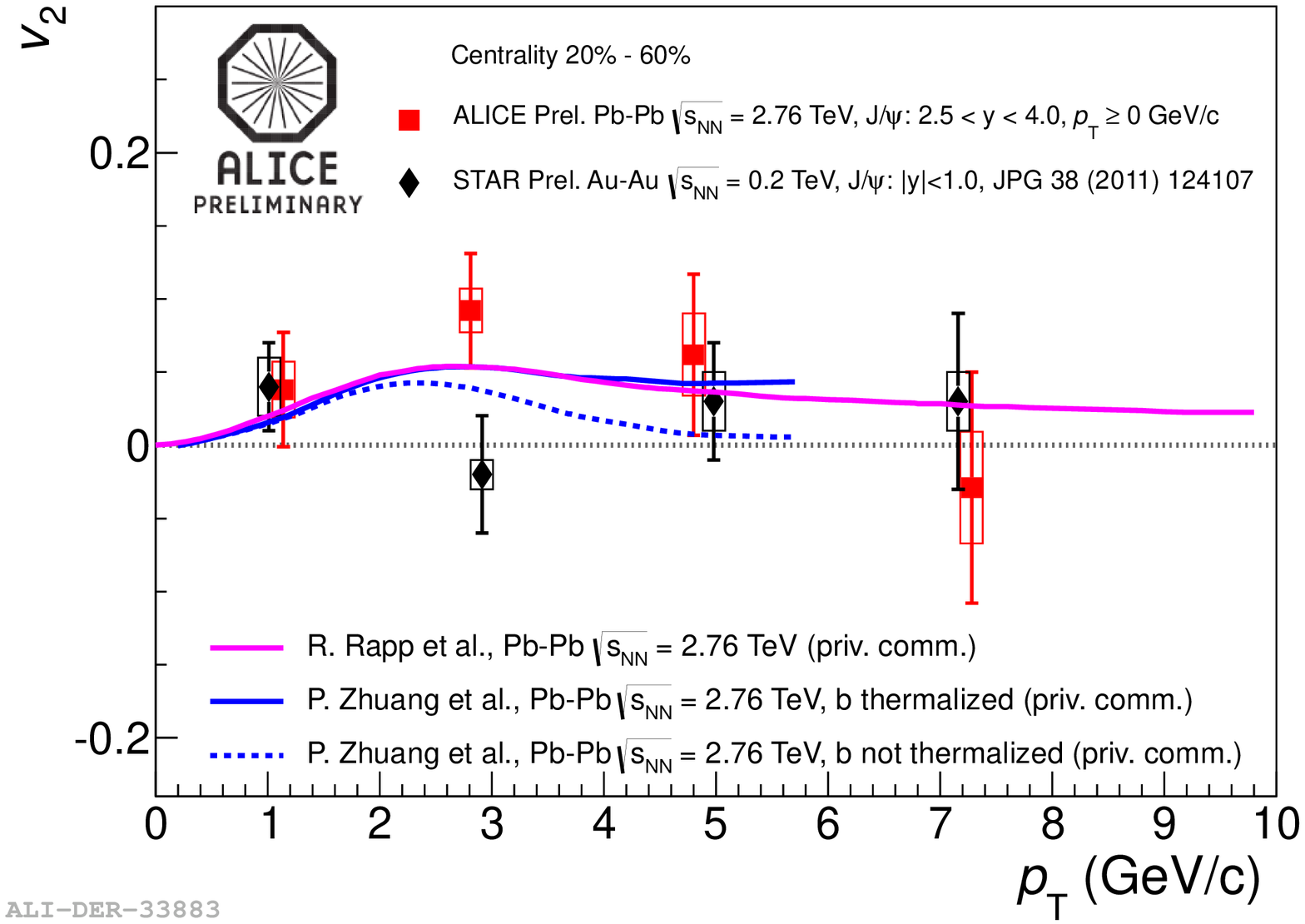}
\end{center}
\caption{Left: $v_2$  extraction with the event plane method in which the J/$\psi$ raw yield is plotted as a function of $\Delta\varphi$ using a fit to the data with ${dN^{{\mathrm{J}}/\psi}}/{d\Delta\varphi} = A \times (1 + 2v_{2}^{\mathrm{obs}} \cos 2\Delta\varphi)$. 
Right: Inclusive J/$\psi$ $v_{2}$ in the centrality bin 20\%-60\% as a function of $p_{\mathrm T}$. A
comparison to STAR results and to two parton transport model calculations~\cite{zhuangModel, rappModel} are shown. 
The vertical bars show the statistical uncertainties, and the boxes indicate the point-to-point uncorrelated systematic uncertainties, which are dominated by the signal extraction. }
\label{Fig:EPmethod}
\end{figure}

The J/$\psi$ $v_2$  is measured using event plane based methods~\cite{voloshin}.
To make a direct comparison with lower energy measurements, the	inclusive J/$\psi$ $v_2$ ($p_{\mathrm{T}}$) was calculated in the same centrality range 20\%-60\% as at RHIC~\cite{starJpsiFlow}, as discussed in detail in~\cite{jpsiv2hp2012}. The $v_2$ is extracted by fitting the J/$\psi$ raw yield as a function of $\Delta\varphi = \phi_{\mathrm {dimuon}} - \Psi_{{\mathrm {EP}}, 2}$ with 
${dN^{{\mathrm{J}}/\psi}}/{d\Delta\varphi} = A\times (1 + 2v_2^{\mathrm{obs}} \cos 2\Delta\varphi)$,
where $A$ is a normalization constant (the standard event plane method), as shown in Fig.~\ref{Fig:EPmethod} (left panel). At LHC, the event-plane-resolution-corrected $v_2$  of J/$\psi$ with $2<p_{\mathrm{T}}<4$~GeV/$c$ is different from the STAR preliminary measurement which is compatible with zero in all the measured $p_{\mathrm{T}}$ range, as shown in Fig.~\ref{Fig:EPmethod} (right panel). Two model calculations based on transport mechanism which include a J/$\psi$ regeneration component from deconfined charm quarks in the medium~\cite{zhuangModel, rappModel} are compared with data. These two models differ mostly in the rate equation controlling the J/$\psi$ dissociation and regeneration. In both models about 50\% of the produced J/$\psi$ mesons originate
from regeneration in QGP in the most central collisions. On one hand, thermalized charm quarks in the medium will transfer a significant elliptic flow to regenerated J/$\psi$. The maximum $v_2$  at $p_{\mathrm{T}} \approx 2.5$~GeV/$c$ results from a dominant contribution of regeneration at lower $p_{\mathrm{T}}$ with respect to the initial J/$\psi$ component. Both models are able to qualitatively describe J/$\psi$ $v_{2} (p_{\mathrm{T}})$ data as both were also able to describe the earlier $R_{\mathrm{AA}}$  measurement~\cite{aliceJpsiPRL}.

\begin{figure}[htbp]
\begin{center}
\includegraphics[width=0.45\linewidth,keepaspectratio]{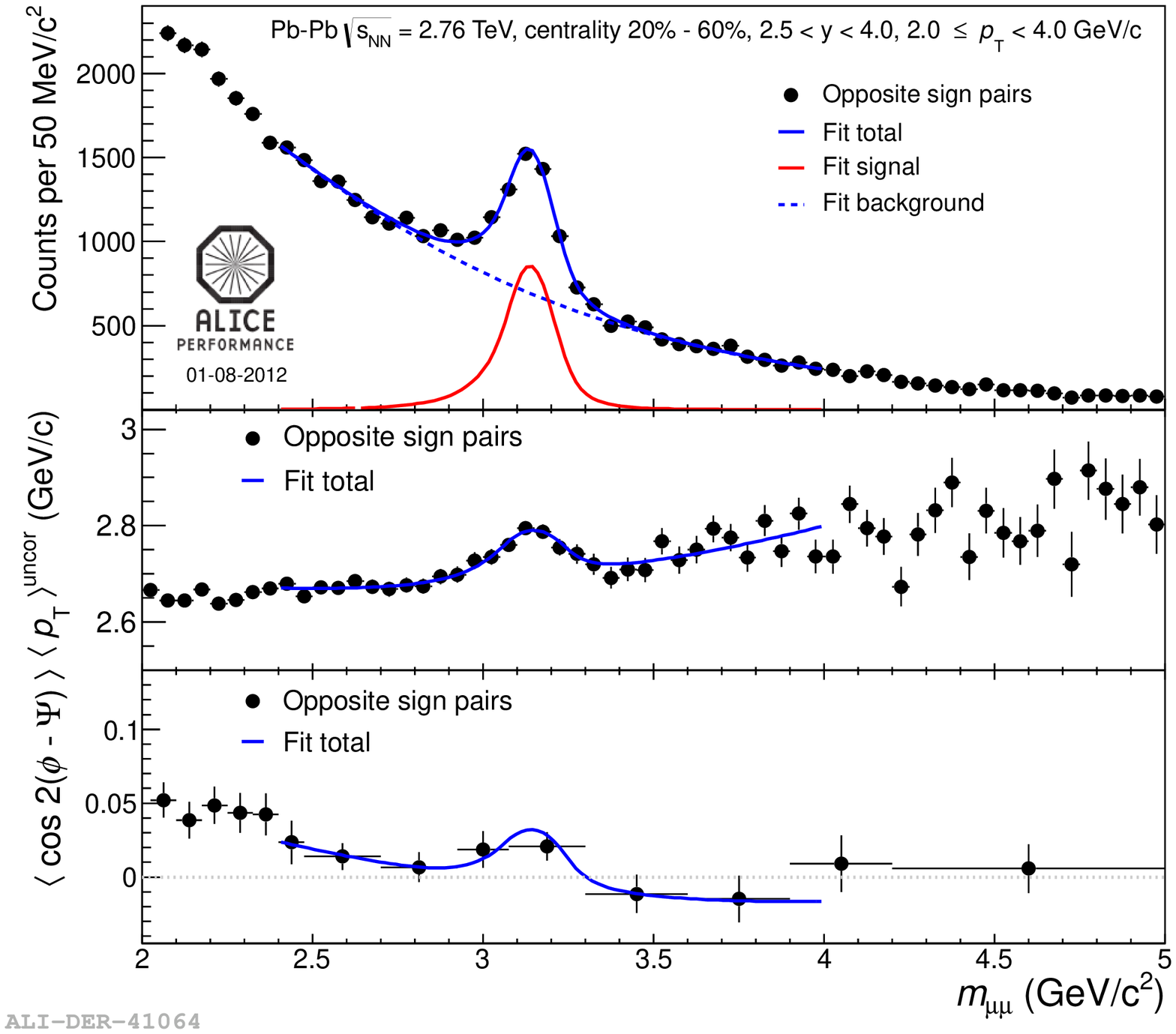} 
\includegraphics[width=0.54\linewidth,keepaspectratio]{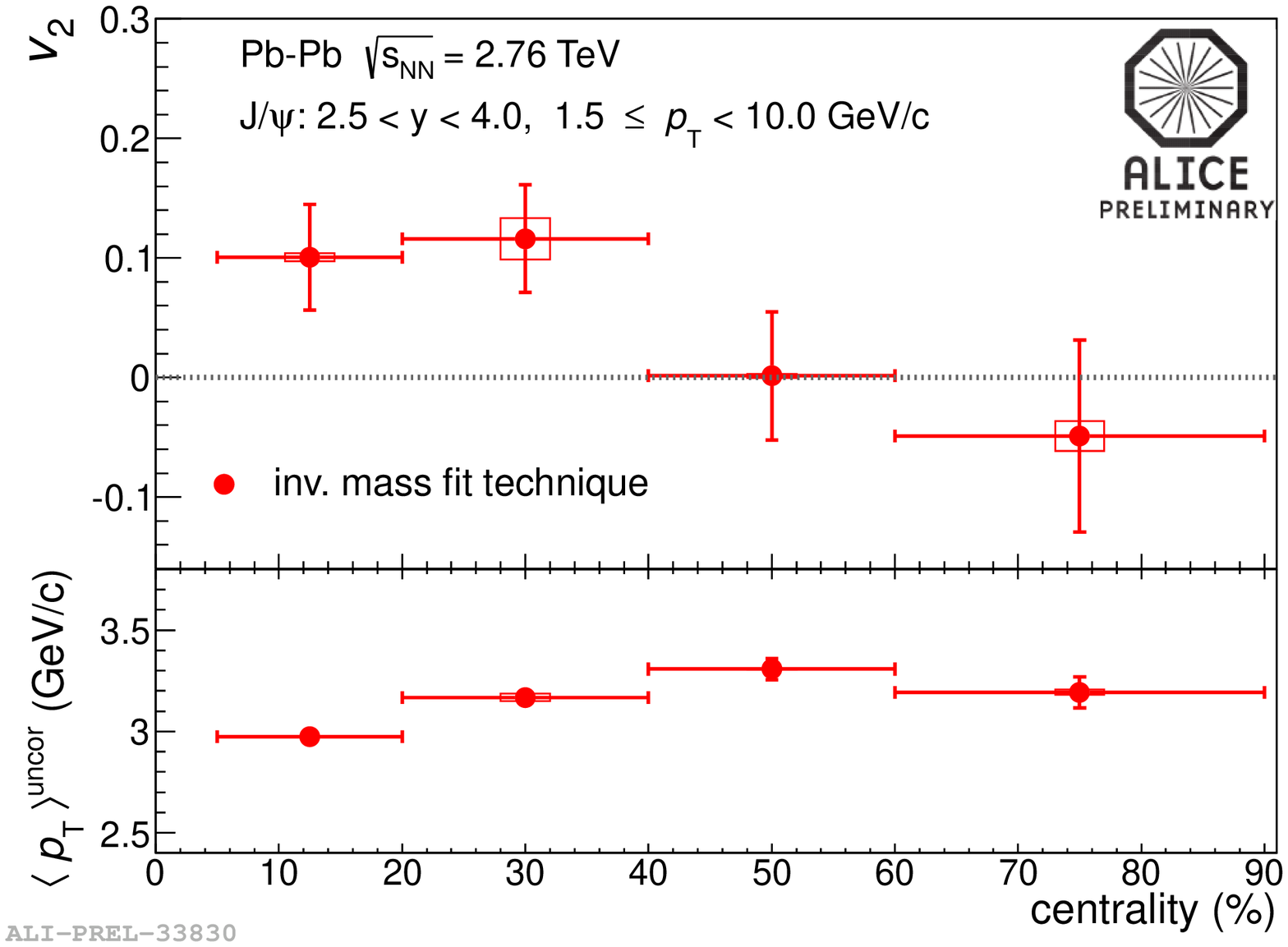}                              
\end{center}
\caption{Left: $\langle p_{T}\rangle^{\mathrm{uncor}}$ and $v_2$  extraction with the fit invariant mass technique. Right: Event plane resolution corrected  J/$\psi$ $v_2$  as a function of centrality of J/$\psi$ with $p_{\mathrm{T}} \geq$ 1.5~GeV/$c$. The vertical bars show the statistical uncertainties, and the boxes indicate the point-to-point uncorrelated systematic uncertainties, which are dominated by the signal extraction.}
\label{Fig:invmassfit}
\end{figure}

Consistent results in the same centrality bin are obtained with an invariant mass fit technique, in which we fit the $v_{2} = \langle \cos 2\Delta\varphi \rangle$ vs. invariant mass ($m_{\mu\mu}$) as described in~\cite{flowMethod}.  The method involves calculating the $v_2$ of the OS dimuons as a function of $m_{\mu\mu}$ and then fitting the resulting $v_2$  ($m_{\mu\mu}$) distribution using:
$v_{2}(m_{\mu\mu}) = v_{2}^{\mathrm{sig}} \alpha(m_{\mu\mu}) + v_{2}^{\mathrm{bkg}}(m_{\mu\mu}) [1-\alpha(m_{\mu\mu})]$, 
where $v_2^{\mathrm{sig}}$ is the J/$\psi$ elliptic flow and $v_2^{\mathrm{bkg}}$ is the background flow (parametrized using a second order polynomial function in this analysis). $\alpha(m_{\mu\mu}) = S/(S + B)$ is the ratio of the signal over the sum of the signal plus background of the $m_{\mu\mu}$ distributions. 
$\alpha(m_{\mu\mu})$ is extracted from fits to the OS invariant mass distribution in each $p_{\mathrm{T}}$ and centrality class. 
The OS dimuon invariant mass distribution was fitted with a Crystal Ball (CB) function to reproduce the J/$\psi$ line shape, and either a third order polynomial or a variable width gaussian to describe the underlying continuum. 
The CB function connects a Gaussian core with a power-law tail~\cite{jpsiShape} at low mass to account for energy loss fluctuations and radiative decays. 
The combination of several CB and underlying continuum parametrization were tested to assess the signal and the related systematic uncertainties. 
The J/$\psi$ $v_2$  in each $p_{\mathrm{T}}$ and centrality class was determined as the average of the $v_2^{\mathrm{sig}}$ obtained by fitting $v_2 (m_{\mu\mu})$ 
with various background shapes, 
while the corresponding systematic uncertainties were defined as the {\it r.m.s.} of these results. A similar method is used to extract the uncorrected average transverse momentum $\langle p_{\mathrm{T}} \rangle^{\mathrm{uncor}}$ of the reconstructed J/$\psi$ in each centrality and $p_{\mathrm{T}}$ class. 
Fig.~\ref{Fig:invmassfit} (left panel) shows typical fits of the OS invariant mass distribution (top left), the $\langle \cos 2(\phi - \Psi_{\mathrm{EP}, 2})\rangle$ (bottom left) and $\langle p_{\mathrm T} \rangle^{\mathrm{uncor}}$ (middle left) as a function of $m_{\mu\mu}$ in the 20\%-60\% centrality class. The obtained J/$\psi$ $\langle p_{\mathrm{T}}\rangle^{\mathrm{uncor}}$ is used to locate the ALICE points when plotted as a function of transverse momentum. 

Fig.~\ref{Fig:invmassfit} (top right) shows $v_2$ for inclusive J/$\psi$ with $p_{\mathrm{T}} \geq 1.5$~GeV/$c$ as a function of centrality. 
The vertical bars show the statistical uncertainties while the boxes indicate the point-to-point uncorrelated systematic uncertainties from the signal extraction. 
The measured $v_2$  depends on the $p_{\mathrm{T}}$ distribution of the reconstructed J/$\psi$. Therefore, $\langle p_{\mathrm{T}}\rangle^{\mathrm{uncor}}$ of the reconstructed $v_2$  is also shown in Fig.~\ref{Fig:invmassfit} (bottom right) as a function of centrality. 
For the two most central bins, 5\%-20\% and 20\%-40\% the inclusive J/$\psi$ $v_2$  for $p_{\mathrm{T}} \geq 1.5$~GeV/$c$ are $0.101 \pm 0.044({\mathrm{stat.}}) \pm 0.003({\mathrm{syst.}})$ and $0.116 \pm 0.045({\mathrm{stat.}}) \pm 0.017({\mathrm{syst.}})$, respectively. 
For the two most peripheral bins the $v_2$  is consistent with zero within uncertainties. 
Although there is a small variation with centrality, the $\langle p_{T}\rangle^{\mathrm{uncor}}$ stays in the range (3.0, 3.3)~GeV/$c$ indicating that the bulk of the reconstructed J/$\psi$ are in the same intermediate $p_{\mathrm{T}}$ range for all centralities. Thus, the observed centrality dependence of the $v_2$  for inclusive J/$\psi$ with $p_{\mathrm{T}} \geq1.5$~GeV/$c$ does not result from any bias in the sampled $p_{\mathrm{T}}$ distributions. 
\begin{figure}[htbp]
\begin{center}
\includegraphics[width=0.55\linewidth,keepaspectratio]{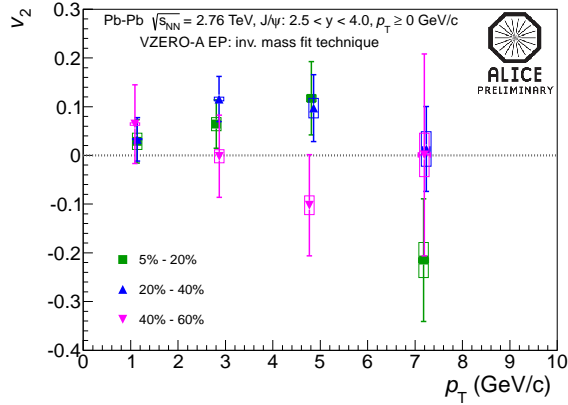}
\end{center}
\caption{J/$\psi$ $v_2$  as a function of $p_{\mathrm{T}}$ in various centrality bins: 5\%-20\%, 20\%-40\% and 40\%-60\%. The vertical bars show the statistical uncertainties, and the boxes indicate the point-to-point uncorrelated systematic uncertainties from the signal extraction.}
\label{Fig:v2cent}
\end{figure}

Fig.~\ref{Fig:v2cent} shows the inclusive J/$\psi$ $v_2 (p_{\mathrm{T}})$, using the invariant mass fit technique,  for central, semi-central and peripheral Pb-Pb collisions at 2.76~TeV. In the semi-central (20\%-40\%) case, taking into account statistical and systematic uncertainties, the combined significance of a non-zero $v_2$ in $2\leq p_{\mathrm{T}} < 6$~GeV/$c$ range is $3\sigma$. At lower and higher transverse momentum the inclusive J/$\psi$ $v_2$  is compatible with zero within uncertainties. In most central (5\%-20\%) and peripheral (40\%-60\%) case, the large uncertainties do not allow any firm conclusion. 

\section{Summary and conclusion}
In summary, we reported the ALICE measurement of inclusive J/$\psi$ $v_2$ at forward rapidity in Pb-Pb collisions at $\sqrt{s_{_{\mathrm{NN}}}}=2.76$~TeV. For non-central (20\%-60\%) collisions a hint of a non-zero J/$\psi$ elliptic flow is observed in the intermediate $p_{\mathrm{T}}$ range in contrast to the zero $v_2$  observed at RHIC. Indication of a non-zero J/$\psi$ $v_2$ is also observed in semi-central (20\%-40\%) collisions at intermediate $p_{\mathrm{T}}$. 
The integrated $v_2$ of J/$\psi$ with $p_{\mathrm{T}}>1.5$~GeV/$c$ in 5\%-40\% collisions also shows a non-zero behavior. 
These measurements complement our earlier results on J/$\psi$ suppression, where a smaller suppression was seen at low transverse momentum at the LHC compared to RHIC~\cite{aliceJpsiPRL,qmJpsiTalk,qmDQTalk}. Both results taken together could indicate that a significant fraction of the observed J/$\psi$ are produced from a (re)combination of the initially
produced charm quarks. 
Our J/$\psi$ elliptic flow results in Pb-Pb collisions at $\sqrt{s_{_{\mathrm{NN}}}}=2.76$~TeV are in qualitative agreement with transport models that are able to reproduce our J/$\psi$ $R_{\mathrm {AA}}$ measurement.



\begin{thebibliography}{99}
 \bibitem{satz}T.~Matsui and H.~Satz, 
Phys.~Lett.,~B178, p.~416, 1986.

\bibitem{aliceJpsiPRL} B.~Abelev {\em et~al.} [ALICE Collaboration] 
 Phys.~Rev.~Lett.~109, 072301, 2012.

\bibitem{spsJpsi}
B.~Alessandro {\em et~al.}, 
Eur.~Phys.~J.,~C39, p.~335--345, 2005.

\bibitem{phenixJpsi}
A.~Adare {\em et~al.}, 
Phys.~Rev.~Lett.,~98, p.~232301, 2007;
Phys.~Rev.,~C84, p.~054912, 2011.

 \bibitem{starJpsiFlow} Z.~Tang [STAR collaboration], 
 J.~Phys.~G38, 12417, 2011. 

\bibitem{aliceJINST} 
K.~Aamodt {\em et~al.}, [ALICE Collaboration], JINST 3, S08002, 2008.

 \bibitem{qmJpsiTalk} R.~Arnardi, these proceedings.
 
 \bibitem{voloshin}
 A.~M.~Poskanzer and S.~A.~Voloshin
Phys.~Rev.~C 58, 1671, 1998.

\bibitem{jpsiv2hp2012} L.~Massacrier, Hard Probes 2012 proceedings arXiv:1208.5401, and references therein.

 \bibitem{zhuangModel}Y.-P.~Liu, {\em et~al.}, 
 Phys.~Lett.~B678, pp.~72--76, 2009 and priv. comm. in 2012.

 \bibitem{rappModel}X.~Zhao and R.~Rapp, 
 Nucl.~Phys.,~A 859, pp.~114--125, 2011, R.~Rapp these proceedings, and priv. comm. in 2012.

\bibitem{flowMethod}
N.~Borghini and J.~Ollitrault, Phys.~Rev.~C70, 064905, 2004, arXiv:nucl-th/0407041 [nucl-th].

\bibitem{jpsiShape}
J.~E.~Gaiser, 
Ph.D. thesis, Standford (1982), appendix-F, SLAC-R-255. 

\bibitem{qmDQTalk} E.~Scomparin, these proceedings.

\end{thebibliography}
\end{document}